\documentstyle[12pt,epsf]{article}
\newcommand{\nc}{\newcommand}
\nc{\postscript}[2] 
{\setlength{\epsfxsize}{#2\hsize}\centerline{\epsfbox{#1}}}
\nc{\bg}{B. Grzadkowski}
\nc{\non}{\nonumber}
\def\dps{\displaystyle}
\def\mib#1{\mbox{\boldmath $#1$}}

\def\bra#1{\langle #1 |} \def\ket#1{|#1\rangle}
\def\vev#1{\langle #1\rangle}
\def\sst#1{\scriptscriptstyle{#1}}

\nc{\barx}{\bar{x}}\nc{\pbarn}{\;\hbox {pb}}\nc{\fbarn}{\;\hbox {fb}}
\nc{\hc}{\hbox {h.c.}} \nc{\re}{\hbox {Re}} 
\nc{\mev}{\hbox {MeV}} \nc{\gev}{\;\hbox {GeV}}
\def\gesim{\lower0.5ex\hbox{$\:\buildrel >\over\sim\:$}} 
\def\lesim{\lower0.5ex\hbox{$\:\buildrel <\over\sim\:$}} 
\nc{\prd}[3]{{\it Phys.\ Rev.}\ {{\bf D{#1}} (#2), #3}}
\nc{\prl}[3]{{\it Phys.\ Rev.\ Lett.}\ {{\bf {#1}} (#2), #3}}
\nc{\plb}[3]{{\it Phys.\ Lett.}\ {{\bf B{#1}} (#2), #3}}
\nc{\npb}[3]{{\it Nucl.\ Phys.}\ {{\bf B{#1}} (#2), #3}}
\nc{\ptp}[3]{{\it Prog.\ Theor.\ Phys.}\ {{\bf {#1}} (#2), #3}}
\nc{\zfp}[3]{{\it Z.\ Phys.}\ {{\bf C{#1}} (#2), #3}}
\nc{\mpla}[3]{{\it Mod.\ Phys.\ Lett.}\ {{\bf A{#1}} (#2), #3}}
\nc{\rmp}[3]{{\it Rev.\ Mod.\ Phys.}\ {{\bf {#1}} (#2), #3}}
\nc{\ijmpa}[3]{{\it Int.\ J.\ of\ Mod.\ Phys.}\
               {{\bf A{#1}} (#2), #3}}
\nc{\epj}[3]{{\it Eur.\ Phys.\ J.}\ {{\bf C{#1}}  (#2), #3}}
\nc{\ttbar}{t\bar{t}}         \nc{\bbbar}{b\bar{b}}
\nc{\tanb}{\tan \beta}        \nc{\twbdec}{t\to W^+ b}
\nc{\tbwbdec}{\bar{t}\to W^- \bar{b}}
\nc{\epem}{e^+e^-}            \nc{\eett}{\epem \to \ttbar}
\nc{\sigeett}{\sigma_{e\bar{e}\to\ttbar}}
\nc{\wpwm}{W^+W^-}            \nc{\tbar}{\bar{t}}
\nc{\bbar}{\bar{b}}           \nc{\wpp}{W^+}
\nc{\mt}{m_t}    \nc{\mts}{m_t^2}   \nc{\mw}{m_W}    \nc{\mws}{m_W^2}
\nc{\mz}{m_Z}    \nc{\mzs}{m_Z^2}
\nc{\ttbardec}{\ttbar \to W^+W^-\bbbar}
\nc{\wwbb}{W^+W^-\bbbar}      \nc{\sm}{SM}
\nc{\cw}{\cos\theta_W}        \nc{\sw}{\sin\theta_W}
\nc{\sws}{\sin^2\theta_W}     \nc{\sig}{\sigma_{tot}}
\nc{\lp}{\ell^+}              \nc{\lm}{\ell^-}
\nc{\epsl}{\epsilon_L}        \nc{\cp}{C\!P}

\nc{\splus}{s_+}       \nc{\smin}{s_-}        \nc{\eps}{\epsilon}
\nc{\psp}{Ps_+}        \nc{\psm}{Ps_-}        \nc{\lsp}{ls_+}
\nc{\lsm}{ls_-}        \nc{\sss}{s_+s_-}      \nc{\m}{m_t}
\nc{\mq}{m_t^2}        \nc{\mr}{\frac{1}{\m}} \nc{\av}{A_{\gamma}}
\nc{\bv}{B_{\gamma}}   \nc{\az}{A_Z}          \nc{\bz}{B_Z}
\nc{\avs}{A_{\gamma}^2}\nc{\azs}{A_Z^2}       \nc{\bzs}{B_Z^2}
\nc{\dav}{\delta \! A_{\gamma}}   \nc{\dbv}{\delta \! B_{\gamma}}
\nc{\dcv}{\delta C_{\gamma}}      \nc{\ddv}{\delta \! D_{\gamma}}
\nc{\daz}{\delta \! A_Z}          \nc{\dbz}{\delta \! B_Z}
\nc{\dcz}{\delta C_Z}             \nc{\ddz}{\delta \! D_Z}
\nc{\dev}{\delta \! E_{\gamma}}   \nc{\dez}{\delta \! E_Z}
\nc{\dfv}{\delta \! F_{\gamma}}   \nc{\dfz}{\delta \! F_Z}
\nc{\rdav}{{\rm Re}(\delta \! A_{\gamma}) \:}
\nc{\rdbv}{{\rm Re}(\delta \! B_{\gamma}) \:}
\nc{\rdcv}{{\rm Re}(\delta C_{\gamma}) \:}
\nc{\rddv}{{\rm Re}(\delta \! D_{\gamma}) \:}
\nc{\rdaz}{{\rm Re}(\delta \! A_Z) \:}
\nc{\rdbz}{{\rm Re}(\delta \! B_Z) \:}
\nc{\rdcz}{{\rm Re}(\delta C_Z) \:}
\nc{\rddz}{{\rm Re}(\delta \! D_Z) \:}
\nc{\idav}{{\rm Im}(\delta \! A_{\gamma}) \:}
\nc{\idbv}{{\rm Im}(\delta \! B_{\gamma}) \:}
\nc{\idcv}{{\rm Im}(\delta C_{\gamma}) \:}
\nc{\iddv}{{\rm Im}(\delta \! D_{\gamma}) \:}
\nc{\idaz}{{\rm Im}(\delta \! A_Z) \:}
\nc{\idbz}{{\rm Im}(\delta \! B_Z) \:}
\nc{\idcz}{{\rm Im}(\delta C_Z) \:}
\nc{\iddz}{{\rm Im}(\delta \! D_Z) \:}
\nc{\cz}{(1+v_e^2)d\:\!'^2}         \nc{\ci}{v_ed\:\!'}
\nc{\ccz}{v_ed\:\!'^2}              \nc{\cci}{d\:\!'}
\nc{\lspace}{\;\;\;\;\;\;\;\;\;\;}  \nc{\llspace}{\lspace \lspace}

\nc{\beq}{\begin{equation}}   \nc{\eeq}{\end{equation}}
\nc{\bea}{\begin{eqnarray}}   \nc{\eea}{\end{eqnarray}}
\nc{\baa}{\begin{array}}      \nc{\eaa}{\end{array}}
\nc{\bit}{\begin{itemize}}    \nc{\eit}{\end{itemize}}
\nc{\ben}{\begin{enumerate}}  \nc{\een}{\end{enumerate}}
\nc{\bce}{\begin{center}}     \nc{\ece}{\end{center}}
\begin{document}
\pagestyle{empty} \setlength{\footskip}{2.0cm}
\setlength{\oddsidemargin}{0.5cm} \setlength{\evensidemargin}{0.5cm}
\renewcommand{\thepage}{-- \arabic{page} --}
\def\mib#1{\mbox{\boldmath $#1$}}
\def\bra#1{\langle #1 |}      \def\ket#1{|#1\rangle}
\def\vev#1{\langle #1\rangle} \def\dps{\displaystyle}
% -------------------------------------------------------------------
   \def\thebibliography#1{\centerline{REFERENCES}
     \list{[\arabic{enumi}]}{\settowidth\labelwidth{[#1]}\leftmargin
     \labelwidth\advance\leftmargin\labelsep\usecounter{enumi}}
     \def\newblock{\hskip .11em plus .33em minus -.07em}\sloppy
     \clubpenalty4000\widowpenalty4000\sfcode`\.=1000\relax}\let
     \endthebibliography=\endlist
   \def\sec#1{\addtocounter{section}{1}\section*{\hspace*{-0.72cm}
     \normalsize\bf\arabic{section}.$\;$#1}\vspace*{-0.3cm}}
% -------------------------------------------------------------------
\vspace*{-1.3cm}
\begin{flushright}
$\vcenter{
% \hbox{IFT-04-00}
\hbox{TOKUSHIMA Report} 
\hbox{(hep-ph/0104105)}
}$
\end{flushright}

\vspace*{0.7cm}
\begin{center}
\renewcommand{\thefootnote}{*}
{\large\bf Probing Anomalous Top-Couplings through the}

\vskip 0.15cm
{\large\bf Final Lepton Angular and Energy Distributions}

\vskip 0.15cm
{\large\bf at Polarized NLC}
\footnote{Talk at {\it Theory Meeting on Physics at Linear Colliders},
March 15 - 17, 2001, KEK, Japan. \\
This work is based on collaboration with \bg.}
\end{center}

\vspace*{0.7cm}
\renewcommand{\thefootnote}{*)}
\begin{center}
{\sc Zenr\=o HIOKI$^{\:}$}\footnote{E-mail address:
\tt hioki@ias.tokushima-u.ac.jp}
\end{center}

\vspace*{0.7cm}
\centerline{\sl Institute of Theoretical Physics,\ 
University of Tokushima}

\vskip 0.14cm
\centerline{\sl Tokushima 770-8502, JAPAN}

\vspace*{2.2cm}
\centerline{ABSTRACT}

\vspace*{0.4cm}
\baselineskip=20pt plus 0.1pt minus 0.1pt
The angular and energy distributions of the final lepton in
$e\bar{e}\to t\bar{t} \to \ell^{\pm} X$ at next linear colliders
(NLC) are analyzed a model-independent way for arbitrary
longitudinal beam polarizations as sensitive tests of possible
anomalous top-quark couplings. The angular-energy distribution
is expressed as a combination of independent functions of the
angle and the energy, where each anomalous parameter is the
coefficient of an individual term. Every parameter could be
thereby determined simultaneously via the optimal-observable
procedure. On the other hand, anomalous $tbW$ couplings totally
decouple from the angular distribution, which enables us to study
$t\bar{t}\gamma/Z$ couplings exclusively.

\vfill
\newpage
%--------------------------------------------------------------------
\renewcommand{\thefootnote}{\sharp\arabic{footnote}}
%--------------------------------------------------------------------
\pagestyle{plain} \setcounter{footnote}{0}
\baselineskip=21.0pt plus 0.2pt minus 0.1pt

% 1111111111111111111111111111111111111111111111111111111111111111111
\sec{Introduction}

The discovery of the top-quark \cite{top} completed the fermion list
required in the standard EW theory (SM). However it is still an open
question whether this quark interacts with the others the standard
way or there exists any new-physics contribution to its couplings. It
decays immediately after being produced because of the huge mass.
Therefore this process is not influenced by any hadronization effects
and consequently the decay products are expected to tell us a lot
about parent top property. 

Next linear colliders (NLC) of $e\bar{e}$ will give us fruitful data
on the top through $e\bar{e}\to t\bar{t}$. In particular the final
lepton(s) produced in its semileptonic decay(s) turns out to carry
useful information of the top-quark couplings \cite{SP}. Indeed many
authors have worked on this subject (see the reference list of
Ref.\cite{GH}), and we also have tackled them over the past several
years.

Here I would like to show some of the results of our latest
model-independent analyses of the lepton distributions for arbitrary
longitudinal beam polarizations \cite{GH}, where we have assumed the
most general anomalous couplings both in the production and decay
vertices in contrast to most of the existing works.\footnote{I would
    like to focus on the final lepton although we studied also the
        $b$ distributions in $e\bar{e}\to t\bar{t} \to b(\bar{b})X$ in
        \cite{GH}.}\ 
What we actually studied are the lepton angular-energy distribution
and the angular distribution, both of which would enable us to
perform interesting tests of the top-quark couplings.

% 2222222222222222222222222222222222222222222222222222222222222222222
\sec{Framework}

We can represent the most general covariant $\ttbar$ couplings to the
photon and $Z$ boson as$\,$\footnote{Throughout this report, I use
    simplified expressions. Here, for example, $A_v$ means $A_v +
        \delta\!A_v$ in our original papers.}
\begin{equation}
{\mit\Gamma}_{vt\bar{t}}^{\mu}=
\frac{g}{2}\,\bar{u}(p_t)\,
\Bigl[\,\gamma^\mu ( A_v-B_v \gamma_5 )
+\frac{(p_t-p_{\bar{t}})^\mu}{2m_t}(C_v-D_v\gamma_5)
\,\Bigr]\,v(p_{\bar{t}})  \label{ff}
\end{equation}
in the $m_e=0$ limit, where $g$ denotes the $SU(2)$ gauge coupling
constant, $v=\gamma,Z$ and
\begin{eqnarray*}
&&\av=4\sw/3,\ \ \bv=0,\ \ \az=(1-8\sin^2\theta_W/3)/(2\cw),\\
&&\bz=1/(2\cw),\ \ C_{\gamma}=D_{\gamma}=C_Z=D_Z=0
\end{eqnarray*}
within the SM. Among the above form factors, $A_{\gamma,Z}$,
$B_{\gamma,Z}$ and $C_{\gamma,Z}$ are parameterizing
$C\!P$-conserving interactions, while $D_{\gamma,Z}$ is
$C\!P$-violating one. 

On the other hand, we adopted the following parameterization of
the $tbW$ vertex suitable for the $\twbdec$ and $\tbwbdec$ decays:
\begin{eqnarray}
&&\!\!{\mit\Gamma}^{\mu}_{Wtb}=-{g\over\sqrt{2}}\:
\bar{u}(p_b)\biggl[\,\gamma^{\mu}(f_1^L P_L +f_1^R P_R)
-{{i\sigma^{\mu\nu}k_{\nu}}\over M_W}
(f_2^L P_L +f_2^R P_R)\,\biggr]u(p_t),\ \ \ \ \ \ \label{ffdef}\\
&&\!\!\bar{\mit\Gamma}^{\mu}_{Wtb}=-{g\over\sqrt{2}}\:
\bar{v}(p_{\bar{t}})
\biggl[\,\gamma^{\mu}(\bar{f}_1^L P_L +\bar{f}_1^R P_R)
-{{i\sigma^{\mu\nu}k_{\nu}}\over M_W}
(\bar{f}_2^L P_L +\bar{f}_2^R P_R)\,\biggr]v(p_{\bar{b}}),
\label{ffbdef}
\end{eqnarray}
where $k$ is the $W$-boson momentum, $P_{L/R}=(1\mp\gamma_5)/2$
and
\[
f_1^L=\bar{f}_1^L=1,\ \ \ 
f_1^R=\bar{f}_1^R=f_2^{L,R}=\bar{f}_2^{L,R}=0
\]
within the SM.
This is also the most general form as long as we treat $W$ as an
on-mass-shell particle, which is indeed a good approximation.
It is worth mentioning that these form factors
satisfy the following relations \cite{cprelation}:
\begin{equation}
f_1^{L,R}=\pm\bar{f}_1^{L,R},\lspace f_2^{L,R}=\pm\bar{f}_2^{R,L},
\label{cprel}
\end{equation}
where upper (lower) signs are those for $C\!P$-conserving
(-violating) contributions.

For the initial beam-polarization we used the following convention:
\begin{equation}
P_{e^{\mp}}=\pm[N(e^{\mp},+1)-N(e^{\mp},-1)]
              /[N(e^{\mp},+1)+N(e^{\mp},-1)],
\end{equation}
where $N(e^{\pm},h)$ is the number of $e^{\pm}$ with helicity $h$
in each beam. Note that $P_{e^+}$ is defined with $+$ sign instead
of $-$ in some literature.

% 3333333333333333333333333333333333333333333333333333333333333333333
\sec{Angular-energy distributions}

After some calculations, we arrived at the following angular-energy
distribution of the final lepton $\ell^+$:
\begin{eqnarray}
&&\frac{d^2\sigma}{dx d\cos\theta}
=\frac{3\pi\beta\alpha^2}{2s} B \:
\Bigl[\:S^{(0)}(x, \theta)                        \non\\
&&\phantom{\frac{d^2\sigma}{dx d\cos\theta}
  =\frac{3\pi\beta\alpha^2}{2s} B\Bigl[}
+\!\!\sum_{v=\gamma,Z}\bigl[\:
 {\rm Re}(\delta\!A_v){\cal F}_{Av}(x, \theta)
+{\rm Re}(\delta\!B_v){\cal F}_{Bv}(x, \theta)    \non\\
&&\phantom{\frac{d^2\sigma}{dx d\cos\theta}
  =\frac{3\pi\beta\alpha^2}{2s} B\Bigl[
  \!\!\sum_{v=\gamma,Z}\Bigl[\,}
+{\rm Re}(\delta  C_v){\cal F}_{Cv}(x, \theta)
+{\rm Re}(\delta\!D_v){\cal F}_{Dv}(x, \theta)
\:\bigr]                                          \non\\
&&\phantom{\frac{d^2\sigma}{dx d\cos\theta}
  =\frac{3\pi\beta\alpha^2}{2s} B\Bigl[}
+{\rm Re}(f^R_2){\cal F}_{2R}(x, \theta)\:\Bigr], \label{A-E}
\end{eqnarray}
where $\beta\:(\equiv\sqrt{1-4m_t^2/s})$ is the top-quark velocity,
$B$ denotes the appropriate branching fraction (=0.22 for $e/\mu$),
$x$ means the normalized energy of $\ell$ defined in terms of its
energy $E$ as
\[
x \equiv \frac{2E}{m_t}\sqrt{\frac{1-\beta}{1+\beta}},
\]
$\theta$ is the angle between the $e^-$ beam direction and the $\ell$
momentum (Fig.\ref{angle}), all in the $\epem$ CM frame, $S^{(0)}$ is
the SM contribution, $\delta\!A_v \sim \delta\!D_v$ express non-SM
part of $A_v \sim D_v$ (i.e., $\delta\!B_{\gamma}$, $\delta C_v$ and
$\delta\!D_v$ are equal to $B_{\gamma}$, $C_v$ and $D_v$ respectively),
and ${\cal F}$ are analytically-expressed functions of $x$ and
$\theta$, which are independent of each other. We neglected all
$|\mbox{non-SM term}|^2$ as a reasonable assumption (see \cite{H-O}).
Replacing $\delta\!D_v$, $f_2^R$ and $\cos\theta$ with $-\delta\!D_v$,
$\bar{f}_2^L$ and $-\cos\theta$ gives the $\ell^-$ distribution.

%\vspace*{0.7cm}
\setlength{\unitlength}{0.7mm} % FFFFFFFFFFFFFFFFFFFFFFFFFFFFFFFFFFFF
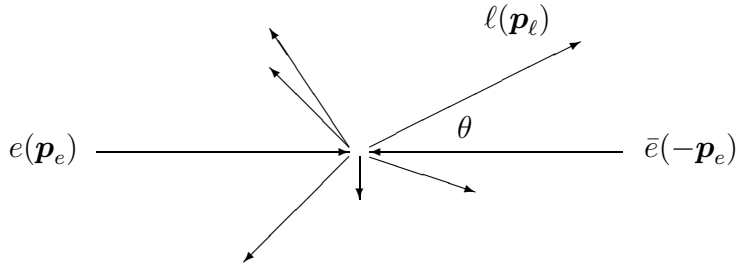
\begin{figure}[b]
\begin{picture}(130,50)(-55,-10)
 \put(0,10){\vector(1,0){48}}
 \put(100,10){\vector(-1,0){48}}
 \put(52,11){\vector(2,1){40}}
\put(48,9){\vector(-1,-1){20}}
\put(50,9){\vector(0,-1){8}}
\put(52,9){\vector(3,-1){20}}
\put(48,11){\vector(-2,3){15}}
\put(48,11){\vector(-1,1){15}}
 \put(-20,5.5){\makebox(20,10){$e(\mib{p}_e)$}}
 \put(103,5.5){\makebox(20,10){$\bar{e}(-\mib{p}_e)$}}
 \put(70,30){\makebox(20,10){$\ell(\mib{p}_{\ell})$}}
% \put(20,-15){\makebox(20,10){$\bar{\mu}(-\mib{p}_\mu)$}}
 \put(60,9.7){\makebox(20,10){$\theta$}}
\end{picture}
\caption{Scattering angle of the final lepton $\ell$}
\label{angle}
\end{figure} % FFFFFFFFFFFFFFFFFFFFFFFFFFFFFFFFFFFFFFFFFFFFFFFFFFFFF

Equation (\ref{A-E}) can be re-expressed as
\begin{equation}
\frac{d^2\sigma}{dx d\cos\theta}
=\frac{3\pi\beta\alpha^2}{2s}B\:
\Bigl[\:{\mit\Theta}_0(x)
+\cos\theta\,{\mit\Theta}_1(x)
+\cos^2\theta\,{\mit\Theta}_2(x) \:\Bigr].
\label{dis1}
\end{equation}
This form directly leads to the angular distribution for $\ell$
through the integration over $x$:
\begin{equation}
\frac{d\sigma}{d\cos\theta}
= \int_{x_-}^{x_+}dx \frac{d^2\sigma}{dx d\cos\theta}=
\frac{3\pi\beta\alpha^2}{2s}B
\left({\mit\Omega}_0+{\mit\Omega}_1
\cos\theta+{\mit\Omega}_2\cos^2\theta\right),   
\label{dis2}
\end{equation}
where ${\mit\Omega}_{0,1,2}
\equiv\int_{x_-}^{x_+}dx\,{\mit\Theta}_{0,1,2}(x)$ and
$x_\pm$ define the kinematical range of $x$.

Surprisingly enough, the non-SM decay part, i.e., $f^R_2$ term completely
disappears through this $x$ integration, and the angular distribution
depends only on the whole production vertex plus the SM decay vertex
\cite{GH,Rin00}.
This never happens in the final $b$-quark distribution.

% 4444444444444444444444444444444444444444444444444444444444444444444
\sec{Numerical analyses}

First, we could determine $\delta\!A_v \sim \delta\!D_v$ and $f^R_2$
simultaneously using the angular-energy distribution (\ref{A-E})
via the optimal-observable procedure \cite{opt}, since these anomalous
parameters are all coefficients of independent functions. In the
second paper of Ref.\cite{GH}, we explored the best $e^{\pm}$
polarizations which minimize the expected statistical uncertainty
($1\sigma$) for each parameter.

Our results assuming the integrated luminosity $L=500\:{\rm fb}^{-1}$
and the lepton-detection efficiency $\epsilon=60\%$ at $\sqrt{s}=500$
GeV are
\begin{equation}
\begin{array}{llllll}
{\mit\Delta}[\:\re(\delta\!A_\gamma)\:]&\!\!=0.16\ &{\rm for}\ 
&P_{e^-}=+0.7\  &{\rm and}\ &P_{e^+}=+0.7, \\
{\mit\Delta}[\:\re(\delta\!A_Z)\:]     &\!\!=0.07\ &{\rm for}\ 
&P_{e^-}=+0.5\  &{\rm and}\ &P_{e^+}=+0.4, \\
{\mit\Delta}[\:\re(\delta\!B_\gamma)\:]&\!\!=0.09\ &{\rm for}\ 
&P_{e^-}=+0.2\  &{\rm and}\ &P_{e^+}=+0.2, \\
{\mit\Delta}[\:\re(\delta\!B_Z)\:]     &\!\!=0.27\ &{\rm for}\
&P_{e^-}=+0.4\  &{\rm and}\ &P_{e^+}=+0.4, \\
{\mit\Delta}[\:\re(\delta C_\gamma)\:] &\!\!=0.11\ &{\rm for}\
&P_{e^-}=+0.1\  &{\rm and}\ &P_{e^+}=\phantom{+}0.0, \\
{\mit\Delta}[\:\re(\delta C_Z)\:]      &\!\!=1.11\ &{\rm for}\
&P_{e^-}=+0.1\  &{\rm and}\ &P_{e^+}=\phantom{+}0.0, \\
{\mit\Delta}[\:\re(\delta\!D_\gamma)\:]&\!\!=0.08\ &{\rm for}\
&P_{e^-}=+0.2\  &{\rm and}\ &P_{e^+}=+0.1, \\
{\mit\Delta}[\:\re(\delta\!D_Z)\:]     &\!\!=14.4\ &{\rm for}\
&P_{e^-}=+0.2\  &{\rm and}\ &P_{e^+}=+0.1, \\
{\mit\Delta}[\:\re(f_2^R)\:]           &\!\!=0.01\ &{\rm for}\
&P_{e^-}=-0.8\ &{\rm and}\ &P_{e^+}=-0.8. \\
\end{array}
\end{equation}
In spite of the large $L$, the precision does not seem so good.
At first sight, readers might conclude that these results
contradict, e.g., the results in \cite{Frey} which give higher
precision. It is however premature to draw such a conclusion.
In \cite{Frey}, they varied just one parameter in one trial,
while we varied all the parameters simultaneously, which is
realistic if we have no other information. I confirmed that
we get a result similar to theirs if we carry out an analysis
the same way. This means that our results could be improved
by any other statistically-independent data.

On the other hand, we can perform another interesting test via
the angular distribution. That is, asymmetries like
\begin{equation}
{\cal A}_{\sst{C\!P}}(\theta)= \Big[\:
{\displaystyle \frac{d\sigma^+(\theta)}{d\cos\theta}-
\frac{d\sigma^-(\pi-\theta)}{d\cos\theta}}
\:\Bigr]\Big/\Bigl[\:
{\displaystyle \frac{d\sigma^+(\theta)}{d\cos\theta}+
\frac{d\sigma^-(\pi-\theta)}{d\cos\theta}}
\:\Bigr]
\end{equation}
or
\begin{equation}
{\cal A}_{\sst{C\!P}}= \frac{
{\displaystyle \int_{-c_m}^{0}\!d\cos\theta
 \frac{d\sigma^{+}(\theta)}{d\cos\theta}
 -\int_{0}^{+c_m}\!d\cos\theta \frac{d\sigma^{-}(\theta)}{d\cos\theta}}}
{{\displaystyle \int_{-c_m}^{0}\!d\cos\theta
  \frac{d\sigma^{+}(\theta)}{d\cos\theta}
 +\int_{0}^{+c_m}\!d\cos\theta \frac{d\sigma^{-}(\theta)}{d\cos\theta}}}, 
\end{equation}

\begin{figure} % FFFFFFFFFFFFFFFFFFFFFFFFFFFFFFFFFFFFFFFFFFFFFFFFFFFFFFF               
\postscript{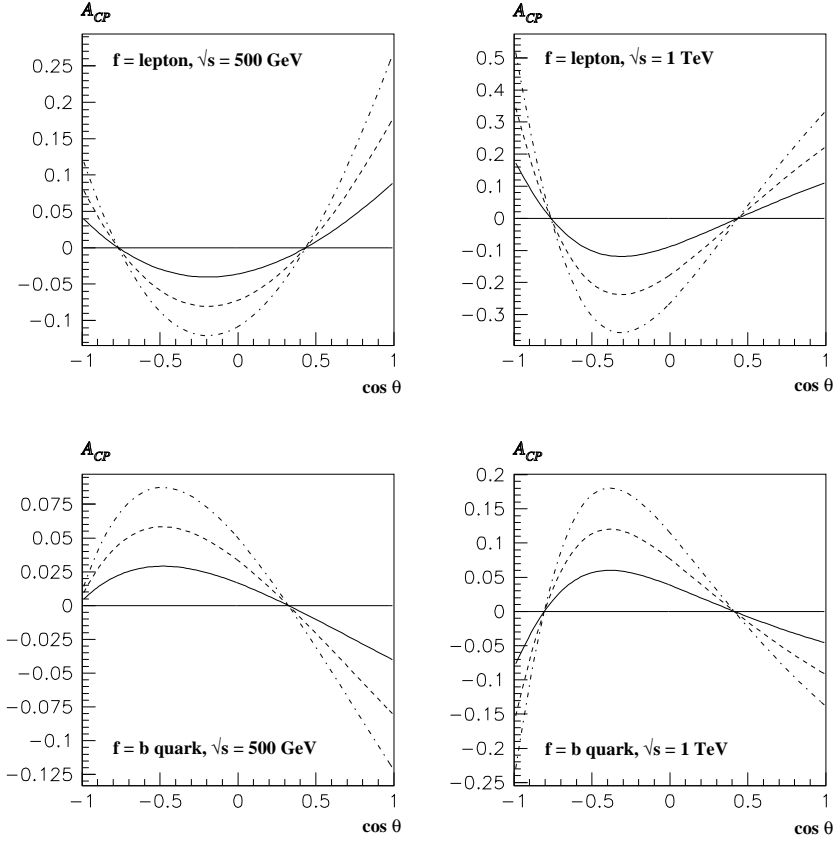}{0.8}
% \vspace*{-0.5cm}    
\caption{$C\!P$-violating asymmetry ${\cal A}_{\sst{C\!P}}(\theta)$ as a
function of $\cos\theta$ for leptonic and $b$-quark distributions for
${\rm Re}(D_\gamma)={\rm Re}(D_Z)={\rm Re}(f_2^R-\bar{f}_2^L)=$0.1
(solid line), 0.2 (dashed line), 0.3 (dash-dotted line) at $\protect
\sqrt{s}=500$ GeV and 1 TeV collider energy. As mentioned in the main
text, ${\cal A}_{\sst{C\!P}}(\theta)$ for lepton only depends on
$D_{\gamma,Z}$.}\label{Th-asym}
\end{figure} % FFFFFFFFFFFFFFFFFFFFFFFFFFFFFFFFFFFFFFFFFFFFFFFFFFFFFFFFF

\noindent
where $d\sigma^{\pm}$ are for $\ell^{\pm}$ respectively and $c_m$
expresses an experimental angle cut, are a pure measure of the
$C\!P$-violating anomalous $t\bar{t}\gamma/Z$ parameters
(Fig.\ref{Th-asym}).

\begin{figure} % FFFFFFFFFFFFFFFFFFFFFFFFFFFFFFFFFFFFFFFFFFFFFFFFFFFFFF
\postscript{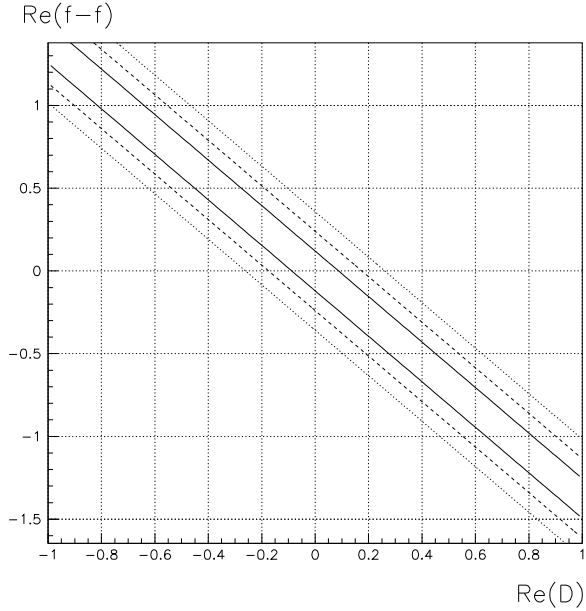}{0.6}
\caption{Parameter area which we can explore through the asymmetry
$A_{\ell\ell}$ introduced in \protect\cite{GH_plb97}. We can confirm
this asymmetry to be non-zero at $1\sigma$, $2\sigma$ and $3\sigma$
level when the parameters $\re{(D_{\gamma,Z})}$ and
$\re{(f^R_2-\bar{f}^L_2)}$ are outside the two solid lines, dashed
lines and dotted lines respectively. Unfortunately there is some
area where two contributions from the production and decay vertices
cancel each other and we get little information.}\label{E-asym}
% \vspace*{0.5cm}
\end{figure}

In Ref.\cite{GH_plb97}, we introduced the following asymmetry
\begin{equation}
A_{\ell\ell}\equiv
\frac
{\dps\int\!\!\int_{x<\bar{x}}dxd\bar{x}\frac{d^2\sigma}{\dps dxd\bar{x}}
 -\int\!\!\int_{x>\bar{x}}dxd\bar{x}\frac{d^2\sigma}{\dps dxd\bar{x}}}
{\dps\int\!\!\int_{x<\bar{x}}dxd\bar{x}\frac{d^2\sigma}{\dps dxd\bar{x}}
 +\int\!\!\int_{x>\bar{x}}dxd\bar{x}\frac{d^2\sigma}{\dps dxd\bar{x}}}
\end{equation}
using the $\ell^{\pm}$ energy correlation $d^2\sigma/dxd\bar{x}$, where
$x$ and $\bar{x}$ are the normalized energies of $\ell^+$ and $\ell^-$
respectively. Generally this is also an asymmetry very sensitive to
$C\!P$ violation. However, when we have no luck and two contributions
from the production and decay vertices cancel each other, we get little
information as found in Fig.\ref{E-asym}. This comparison lightens the
outstanding feature of ${\cal A}_{\sst{C\!P}}(\theta)$ and
${\cal A}_{\sst{C\!P}}$ more clearly.

% \newpage
 % FFFFFFFFFFFFFFFFFFFFFFFFFFFFFFFFFFFFFFFFFFFFFFFFFFFFFFFF

% 5555555555555555555555555555555555555555555555555555555555555555555
\sec{Summary}

I showed here some results of our latest work on the angular and
energy distributions of the lepton ($e$ or $\mu$) produced in $e\bar{e}
\to t\bar{t} \to \ell^{\pm}X$. There the most general covariant forms
were assumed both for the $t\bar{t}\gamma/Z$ and $tbW$ couplings, which
makes our analysis fully model-independent.

The angular-energy distribution $d^2\sigma/dx d\cos\theta$ could enable
us to determine in principle all the anomalous parameters in the general
$t\bar{t}\gamma/Z$ and $tbW$ couplings simultaneously. Although
extremely high luminosity is required to achieve good precision, it
never means our analysis is impractical. We could get better precision
when we have any other independent information on those anomalous
parameters.

On the other hand, the angular distribution $d\sigma/d\cos\theta$ is
completely free from the non-SM decay vertex. Therefore, once we 
catch any non-trivial signal of non-standard phenomena, it will be
an indication of new-physics effects in $t\bar{t}\gamma/Z$ couplings.
This is quite in contrast to asymmetries using the single or double
energy distributions of $e\bar{e}\to t\bar{t} \to \ell^{\pm}X\:/\:
\ell^+ \ell^-X'$, where cancellation between the production and
decay contributions could occur.

\vspace*{0.6cm}
% RRRRRRRRRRRRRRRRRRRRRRRRRRRRRRRRRRRRRRRRRRRRRRRRRRRRRRRRRRRRRRRRRRR

\end{document}